\documentclass[prl,showpacs,amsmath,twocolumn]{revtex4}  
\newcommand{\beq}{\begin{equation}}
\newcommand{\enq}{\end{equation}}

\usepackage{graphicx}

\begin{document}
\title{Instability Versus Equilibrium Propagation of Laser Beam in Plasma}

\author{Pavel M. Lushnikov$^{1,2}$ and Harvey A. Rose$^1$
}

\affiliation{$^1$Theoretical Division, Los Alamos National
Laboratory,
  MS-B213, Los Alamos, New Mexico, 87545\\
  $^2$ Landau Institute for Theoretical Physics, Kosygin St. 2,
  Moscow, 119334, Russia
}

\email{har@lanl.gov}

\date{
November 19, 2003}

\begin{abstract}
We obtain, for the first time, an analytic theory of the forward
stimulated Brillouin scattering instability of a spatially and
temporally incoherent laser beam, that controls the transition
between statistical equilibrium and non-equilibrium (unstable)
self-focusing regimes of beam propagation. The stability boundary
may be used as a comprehensive guide for inertial confinement
fusion designs. Well into the stable regime, an analytic
expression for the angular diffusion coefficient is obtained,
which provides an essential correction to a geometric optic
approximation for beam propagation.
\end{abstract}

\pacs{ 42.65.Jx 52.38.Hb}

\maketitle

Laser-plasma interaction has both fundamental interest and is
critical for future experiments on inertial confinement fusion
(ICF) at the National Ignition Facility (NIF)\cite{Lindl1995}.
NIF's plasma environment, in the indirect drive approach to ICF,
has hydrodynamic length and time scales of roughly millimeters
and 10 ns respectively, while the laser beams that traverse the
plasma, have a transverse correlation length, $l_c$, of a few
microns, and coherence time $T_c$ of roughly  a few ps. These
microscopic fluctuations induce corresponding small-scale density
fluctuations and one might naively expect that their effect on
beam propagation to be diffusive provided self-focusing is
suppressed by small enough \cite{RoseDuBois1993} $T_c$, $T_c  \ll
l_c/c_s$, with $c_s$ the speed of sound. However, we find that
there is a collective regime of the forward stimulated Brillouin
scattering \cite{SchmittAfeyan1998} (FSBS) instability which
couples the beam to transversely propagating low frequency ion
acoustic waves. The instability has a finite intensity threshold
even for very small $T_c$ and can cause strong non-equilibrium
beam propagation (self-focusing) as a result.

We present for the first time, an analytic theory of the FSBS
threshold in the small $T_c$ regime.
In the stable regime, an analytic expression for the beam angular
diffusion coefficient, $D$, is obtained to lowest order in $T_c$,
which is compared with simulation. $D$ may be used to account for
the effect of otherwise unresolved density fluctuations on beam
propagation in a geometric optic approximation. This would then
be an alternative to a wave propagation code
\cite{StillBergerEtAl2000}, that must resolve the beam's
correlation lengths and time, and therefore is not a practical
tool for exploring the large parameter space of ICF designs.
Knowledge of this FSBS threshold may be used as a comprehensive
guide for ICF designs. The important fundamental conclusion is,
for this FSBS instability regime, that even very small $T_c$ may not prevent significant
self-focusing. It places a previously unknown limit in the large parameter space of
ICF designs.

We assume that the beam's spatial and temporal coherence are linked as
in the induced spatial incoherence
\cite{LehmbergObenschain1983} method, which gives a stochastic
boundary condition at $z=0$ ($z$ is the beam propagation direction )
 for the various Fourier transform components
\cite{comment2}, $\hat E$, of the electric field spatial-temporal
envelope, $E$,
\begin{eqnarray}\label{phik}
\hat E({\bf k },z=0,t)=|\hat E({\bf k})|\exp\Big [ i\phi_{\bf
k}(t)\Big ], \nonumber \\
 \Big \langle \exp i\Big [\phi_{\bf
k}(t)-\phi_{{\bf k}'}(t')\Big ]\Big \rangle =\delta_{{\bf k
k}'}\exp\Big (-|t-t'|/T_c\Big).
\end{eqnarray}
 The amplitudes, $|\hat E({\bf
k})|$, are chosen to mimic that of actual
experiments, as in the idealized "top hat" model of NIF
optics:
\begin{eqnarray}\label{tophat}
|\hat E({\bf k})|=const, \ k<k_m; \ |\hat E({\bf k})|=0, \ k>k_m,
\end{eqnarray}
with $1/l_c\equiv k_m\simeq k_0/(2F)$, $F$ the optic $f/\#$, and the average intensity,
$  \Big \langle I\Big \rangle \equiv \Big \langle |E|^2\Big \rangle
=I_0$ determines the constant. At electron densities, $n_e$, small
compared to critical, $n_c$, and for $F^2\gg 1$, $E$ satisfies \cite{comment3}
\begin{equation}\label{Eeq1}
  \Big (i\frac{\partial}{\partial
  z}+\frac{1}{2k_0}\nabla^2-\frac{k_0}{2}\frac{n_e}{n_c}\rho
  \Big )E=0, \ \nabla=(\frac{\partial}{\partial x},\frac{\partial}{ \partial
  y}).
\end{equation}
$k_0$ is $\simeq$ the laser wavenumber in vacuum.  The relative
density fluctuation, $\rho = \delta n_e/n_e$, absent plasma flow
and thermal fluctuations which are ignored here, propagates
acoustically with speed $c_s$:
\begin{equation}\label{neq1}
 (R_0^{\rho\rho})^{-1} \ln (1+\rho)\equiv \Big (\frac{\partial^2}{\partial
  t^2}+2\tilde\nu\frac{\partial}{\partial t}-c_s^2\nabla^2  \Big )\ln (1+\rho)=c_s^2 \nabla^2 I.
\end{equation}
$\tilde \nu$ is an integral operator whose Fourier transform is
$\nu k c_s$, where $\nu$ is the Landau damping coefficient. $E$
is in thermal units defined so that in equilibrium the standard
$\rho=\exp (-I_0)-1$ is recovered. The physical validity of Eqs.
$(\ref{Eeq1}),(\ref{neq1})$ as a model of self-focusing in plasma
has been discussed before
\cite{KawSchmidtWilcox1973,SchmittOng1983,Schmitt1988}. If
$n_e/n_c$ is taken constant, there are 3 dimensionless parameters
for  $\rho\ll 1$: $\nu$,$ \,  \tilde I_0\equiv
(k_0/k_m)^2(n_e/n_c)I_0/\nu, \,$
 and $\tilde T_c\equiv k_mc_sT_c$.

Since Eqn. $(\ref{Eeq1})$ is linear in $E$, it may be decomposed,
at any $z$, into a finite sum, $E=\sum_{j}E_{{\bf m}_j}({\bf
x},z,t)$, where each term has a typical wavevector ${\bf m}_j:$
$E_{{\bf m}_j}({\bf x},z=0,t)\sim \exp(i{{\bf m}_j}\cdot {\bf
x})$.  Cross terms $E_{{\bf m}_j}E^*_{{\bf m}_{j'}}, \ {\bf
m}_j\neq {{\bf m}_j'}$, in the intensity, vary on the times cale
$\tilde T_c$
 so that their effect on the density response,  Eq. $(\ref{neq1})$,  is suppressed for $\tilde T_c\ll 1$
 (see detailed discussion in
\cite{RoseDuBoisRussell1990}). Similar consideration may be
applied to general media with slow nonlinear response, including
photorefractive media \cite{Segev1997}. Then the rhs of Eq.
$(\ref{neq1})$ can be approximated as
\begin{eqnarray}\label{nFeq1}
  c_s^2 \nabla^2 I=c_s^2 \nabla^2\sum\limits_{j}|E_{{\bf m}_j}|^2=c_s^2 \nabla^2
  \int d{\bf v} F({\bf x},{\bf v},z,t). \\
\label{Fdef1}
 F({\bf x},{\bf v},z,t)=\int d{\bf r}\sum\limits_{j j'}\delta_{{\bf m}_j {\bf m}_{j'}}\nonumber \\
 \times E_{{\bf m}_j}({\bf x}-{\bf r}/2,z,t)E^*_{{\bf m}_{j'}}({\bf
 x}+{\bf r}/2,z,t)e^{i{\bf v}\cdot {\bf r}}/(2\pi)^2
\end{eqnarray}
is a variant of the Wigner distribution function which satisfies,
as follows from Eq. $(\ref{Eeq1})$,
\begin{eqnarray}\label{Fteq1}
  \frac{\partial F}{\partial z} +2{\bf v}\cdot \frac{\partial F}{\partial {\bf
  x}}-\frac{i}{\pi^2}\int\Big[\hat\rho\big(-2[{\bf v}-{\bf v}'],z,t\big)\times \nonumber \\
  \exp\big (-2i[{\bf v}-{\bf v}']\cdot {\bf x}\big )
  -\hat\rho\big(2[{\bf v}-{\bf v}'],z,t\big)\times \nonumber \\
  \exp\big (2i[{\bf v}-{\bf v}']\cdot {\bf x}\big )\Big ]F({\bf x},{\bf
  v}',z,t)d{\bf v}'=0,
\end{eqnarray}
with boundary value $F({\bf x},{\bf v},z=0,t)\equiv F_0({\bf
v})=|\hat E({\bf v})|^2$. Here the unit of $x$ is
$(1/k_0)\sqrt{n_c/n_e}$ and that of $z$ is $(2/k_0)n_c/n_e$.
Zero density fluctuation, $\rho=\partial \rho/\partial t=0,$ is an
equilibrium solution of (4), $(\ref{nFeq1})$ and $(\ref{Fteq1})$,
whose linearization admits solutions of the form, $\delta \rho
\sim e^{\lambda z}\exp i({\bf k}\cdot {\bf x}-\omega t)$, for real
$\bf k$ and $\omega$, with
\begin{eqnarray}\label{lambeq1}
 {\tilde \lambda \equiv}k_0\lambda/k_m^2=\frac{\tilde k(i\tilde I_0-2f)}{2\tilde I_0}
 \left [\frac{f^2\tilde k^2-if\tilde I_0\tilde k^2-\tilde I_0^2}
 {f(f-i\tilde I_0)} \right ]^{1/2}, \nonumber \\
f\equiv \frac{\omega^2-k^2c_s^2+2i\nu\omega kc_s}{2i\nu k^2c_s^2}, \ \tilde k\equiv \frac{k}{k_m}.
\end{eqnarray}
Here and below we assume that the principle branches of square and
cubic roots are always chosen so that the branch cut in the
complex plane is on the negative axis and values of square root
and cubic root are positive for positive values of their
arguments. The real part of $\lambda$, $\lambda_r\equiv
Re(\lambda)$ has a maximum, as a function of $\omega$, close to
resonance, $\omega=\pm k c_s [1+O(\nu)]$. Below we calculate all
quantities at resonance $\omega=\pm k c_s$ because analytical
expressions are much simpler in that case. $\lambda_r(k)$ has a
maximum, $\lambda_{max}=k_m^2\tilde \lambda_{max}/k_0>0$, at $k
\equiv k_{max}$,
\begin{eqnarray}\label{kmax}
k_{max}/k_m= \tilde I_0 \sqrt{7(3\, \tilde I_0^2-2)2^{2/3}c^{-1}+8-2^{1/3}c}\times\nonumber \\
\big [3^{1/2}2(1+\tilde I_0^2)^{1/2}\big ]^{-1}, \nonumber \\
 c=(c_1+c_2)^{1/3}, \quad c_1=-40 +
225 \tilde I_0^2 -27\tilde I_0^4, \nonumber \\ c_2=-3i(\tilde
I_0^2+4)\sqrt{27-60\tilde I_0^2-81\tilde I_0^4},
\end{eqnarray}
  Modes
with $k> k_{cutoff}$ are stable ($\lambda_r<0$), with $k_{cutoff}
=k_m\tilde I_0^2(1+\tilde I_0^2)^{-1/2}/2,$ which defines a
wavenumber-dependent FSBS threshold.

 As $\tilde
I_0 \to 0,$ at fixed $k$, $k_0 \lambda_r \to -k^2/\tilde I_0,$
recovering the $\delta(z)$ behavior of density response function
$R^{\rho\rho}_0$ in $(\ref{neq1})$. If $k_m$ is set to zero, the
coherent forward stimulated Brillouin scattering (FSBS)
convective gain rate \cite{SchmittAfeyan1998} is recovered in the
paraxial wave approximation. Unlike the static response,
$\lambda(k,\omega=0),$ which is stable \cite{comment4} for all
$k$ for small enough $I_0 $, the resonant response remains
unstable at small $k$ \cite{comment5} since as $\tilde I_0\to 0,
\ \tilde \lambda_{max}\to 0.024\tilde I_0^5  $ and $k_{cutoff}\to
k_m \tilde I_0^2/2$.

Since the FSBS instability peaks near $\omega=\pm kc_s,$ one
expects an acoustic-like peak to appear in the intensity
fluctuation power spectrum, $|I(k,\omega)|^2$, for $k$ less than
$k_{cutoff}$ as in the simulation ($f/8,\;{{n_e} \mathord{\left/
{\vphantom {{n_e} {n_c}}} \right. \kern-\nulldelimiterspace}
{n_c}}=0.1$) results shown in figure 1.
\begin{figure}
\begin{center}
\includegraphics[width = 3.4 in]{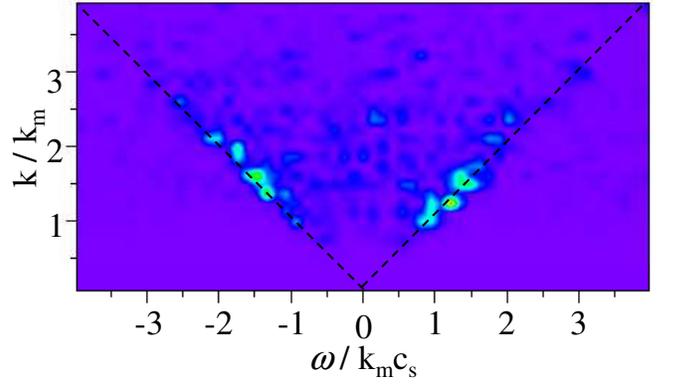}
\caption{Density source power spectrum, $k^4|I(k,\omega)|^2$,
with $\tilde I_0\simeq 4.12, \ \nu=0.15, \tilde T_c\simeq 0.033,$ and
$k_m^2z/k_0\simeq 7.9.$ The dashed lines are at $\omega=\pm kc_s$.
} \label{fig:fig1}
\end{center}
\end{figure}
 The fraction of power in this acoustic peak,
${{\int\limits_{{{2kc_s} \mathord{\left/ {\vphantom {{2kc_s} 3}}
\right. \kern-\nulldelimiterspace} 3}<\left| \omega
\right|<{{4kc_s} \mathord{\left/ {\vphantom {{4kc_s} 3}} \right.
\kern-\nulldelimiterspace} 3}} {\left| {I\left( {k,\omega }
\right)} \right|^2d\omega }} \mathord{\left/ {\vphantom
{{\int\limits_{{{2kc_s} \mathord{\left/ {\vphantom {{2kc_s} 3}}
\right. \kern-\nulldelimiterspace} 3}<\left| \omega
\right|<{{4kc_s} \mathord{\left/ {\vphantom {{4kc_s} 3}} \right.
\kern-\nulldelimiterspace} 3}} {\left| {I\left( {k,\omega }
\right)} \right|^2d\omega }} {\int_{-\infty }^{+\infty } {\left|
{I\left( {k,\omega } \right)} \right|^2d\omega }}}} \right.
\kern-\nulldelimiterspace} {\int_{-\infty }^{+\infty } {\left|
{I\left( {k,\omega } \right)} \right|^2d\omega }}} \qquad,$
 increases
significantly as $\tilde I_0$ passes through its threshold value
for a particular $k$, as shown in figure 2.
\begin{figure}[htbp]
\begin{center}
\includegraphics[width = 3.4 in]{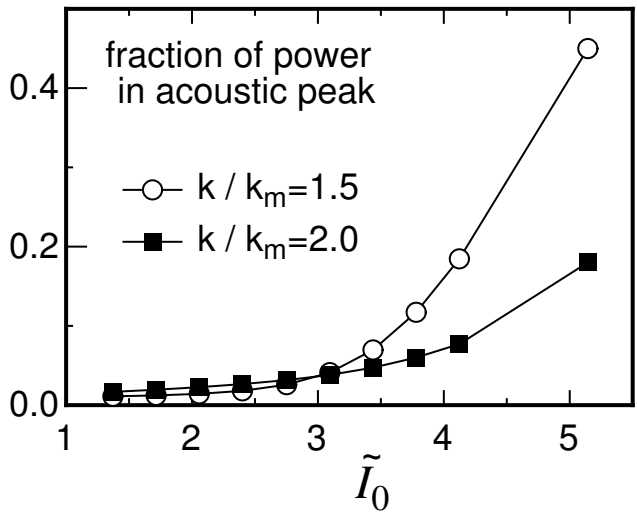}
\caption{Fractional power in acoustic peak of the {\it intensity
fluctuation spectrum}, with parameters as in figure 1, except
$k_m^2z/k_0\simeq 5.2.$ Note that the FSBS intensity threshold
for $k/k_m=$ 1.5 (2.0) is about 3 (4)} \label{fig:fig2}
\end{center}
\end{figure}
There is no discernible difference in shape between
$|E(k,\omega,z)|^2$  at $z=0$, where it is $\propto 1/\big
[1+(\omega T_c)^2\big ]$, and at finite $z$, for small $T_c$.

If $\tilde \lambda _{\max }\ll 1$, i.e., $\tilde I_0\lesssim 1,$ then
the FSBS growth length, ${1 \mathord{\left/ {\vphantom {1 {\lambda
_{\max }}}} \right. \kern-\nulldelimiterspace} {\lambda _{\max
}}}$, is
 large compared to the (vacuum) $z$ correlation length,
 $\propto{{k_0} \mathord{\left/ {\vphantom {{k_0} {k_m^2}}}
 \right. \kern-\nulldelimiterspace} {k_m^2}}$,
and it is found, for small $T_c$, that a quasi-equilibrium is
attained: various low order statistical moments  are roughly
constant over the simulation range once $k_m^2z/k_0\gtrsim 5$, as
seen in figure 3.
\begin{figure}[htbp]
\begin{center}
\includegraphics[width = 3.4 in]{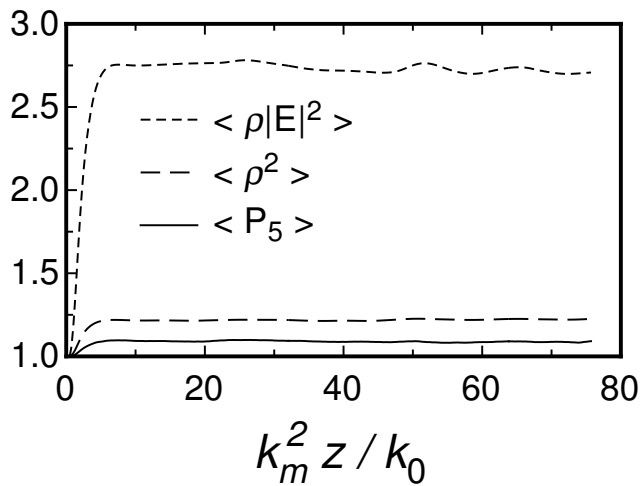}
\caption{A quasi-equilibrium is attained with one point $E$
fluctuations remaining nearly Gaussian, as evidenced by the small
change in $P_5$ \cite{StillBergerEtAl2000}, the fraction of power
with intensity at least $5I_0$, but strongly modified $I-\rho$
correlations. Parameters are $\tilde I_0\simeq 0.53, \, \nu=0.3$
and $\tilde T_c\simeq 0.26$.  Each curve is normalized to its
value at z=0.} \label{fig:fig3}
\end{center}
\end{figure}
A true equilibrium cannot be attained since $\langle k^2\rangle
\equiv \langle|\nabla E|^2\rangle/I_0$ grows due to scattering
from density fluctuations as in figure 4.
\begin{figure}[htbp]
\begin{center}
\includegraphics[width = 3.4 in]{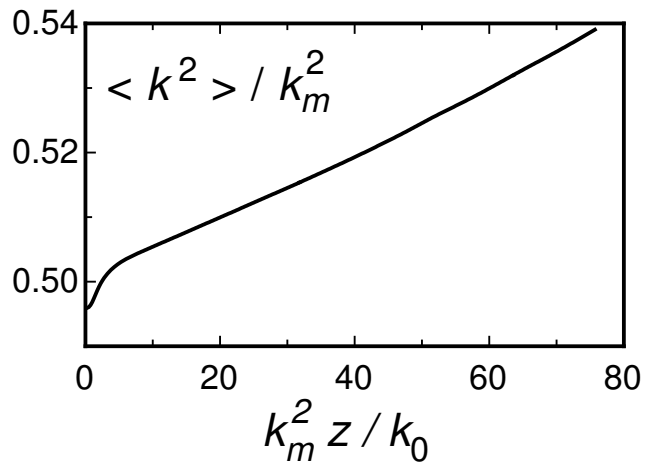}
\caption{For parameters of figure 3, $\langle k^2\rangle \equiv
\langle|\nabla E|^2\rangle/I_0$, increases little over the
 initial equilibration distance of roughly 5 in these units. The
subsequent diffusion rate is 4.4E-04. } \label{fig:fig4}
\end{center}
\end{figure}
 A dimensionless diffusion coefficient, $\tilde D\equiv (k_0 /k_m^4)
\frac{d}{dz}\langle k^2\rangle,$ (proportional to the rate of
angulare diffusion) may be extracted from the data of figure 4 by
fitting a smooth curve to $\langle k^2\rangle$  for $5<k_m^2
z/k_0 <76$, and evaluating its slope, extrapolated to $z=0$. This
yields a diffusion coefficient of 4.4E-04.

$\tilde D$ may be compared to the solution of the stochastic Schroedinger equation (SSE)
 \cite{BalRyzhik2002} with a self-consistent  random potential \cite{Zakharov}, $\rho$,
 whose covariance, $C^{\rho\rho}$ ($C^{\rho\rho}$ is
 a quadratic functional of $F(k)$) is evaluated as follows \cite{Moody2000}.
Take $E$ as given by Eqn. $(\ref{Eeq1})$ with $\rho$ set to $0$
since it goes to zero with $\tilde T_c$, and use it in Eqn.
$(\ref{neq1})$, with $\ln(1+\rho)\to \rho$, to evaluate
$C^{\rho\rho}$. This is consistent only if $\tilde I_0< 1,$ so
that the density responce is stable
 except at small $k/k_m$. It follows, to leading order in $\tilde T_c$,
 that the SSE prediction for $\tilde D,$
for the top hat spectrum,
\begin{equation}\label{Deq1}
 \tilde D_{SSE}=\nu \tilde T_c\tilde I_0^2/68.8\ldots,
\end{equation}
has the value 3.2E-04 for the parameters of Fig. 4.  Note that $\tilde D_{SSE}$ is
proportional to $\langle \rho^2 \rangle$  and the roughly $20\%$ increase of $\langle \rho^2
\rangle$  over its perturbative evaluation (see figure 3) used in the SSE accounts for about
$1/2$ of the difference between $\tilde D$ and $\tilde D_{SSE}$.

We find that $\tilde D$ depends essentially on the spectral form,
$\langle {| {\hat E( k)}|^2} \rangle =F( k )$ , e.g., for
Gaussian $F(k)$ with the same value of $\langle k^2 \rangle$,
$D_{Gaussain}\approx 3D_{top\;hat}$. A numerical example of this
dependence is found in figures 4 and 5.  $\tilde D$  changes by
$40\%$ over $5<k_m^2 z/k_0 <76$, because $F(k)$ changes
significantly as seen in figure 5.
\begin{figure}
\begin{center}
\includegraphics[width = 3.4 in]{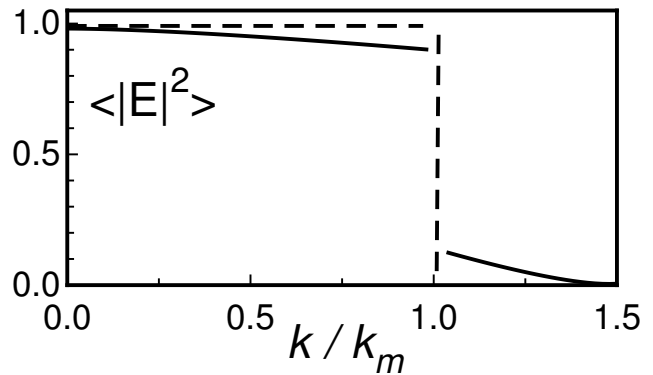}
\caption{Top hat boundary condition, dashed line, changes qualitatively over the
propagation distance shown in figure 4: solid line at $k_m^2z/k_0\simeq 76.$} \label{fig:fig6}
\end{center}
\end{figure}
In this sense, for NIF relevant boundary conditions, angular
diffusion is an essential correction to the geometrical optics
model, which (absent refraction) has constant $F(k)$.

Eqn. (10) implies that ${d \mathord{\left/ {\vphantom {d {dz}}}
\right. \kern-\nulldelimiterspace} {dz}}\left\langle {\left( {{k \mathord{\left/ {\vphantom {k {k_m}}} \right. \kern-\nulldelimiterspace} {k_m}}} \right)^2} \right\rangle \propto {1 \mathord{\left/ {\vphantom {1 {k_m}}} \right. \kern-\nulldelimiterspace} {k_m}}$,
while $\lambda _{\max }\propto {1 \mathord{\left/ {\vphantom {1 {k_m^8}}} \right. \kern-\nulldelimiterspace} {k_m^8}}$.
If the diffusion length is smaller than the FSBS growth length, then propagation, which effectively increases $k_m$, will reinforce this
 ordering.  This stability condition may be expressed as $\tilde D>\tilde \lambda _{\max }$, or qualitatively as \cite{comment6}
\begin{equation}\label{Deq2}
 \nu \tilde T_c>\tilde I_0^3.
\end{equation}
This is a global condition, as opposed to the wavenumber
dependent threshold, $ k_{cutoff}( {\tilde I_0} )$. However, even
if Eqn. (11) is violated, it is not until $k_{cutoff}\approx
1.5k_m$, so  that the peak of the density fluctuation spectrum is
unstable, that FSBS has a strong effect. For these larger $I_0$
values a quasi-equilibrium is not attained,  and it is more
useful to consider an integral measure,  $\triangle (\langle k^2
\rangle,z)\equiv\langle k^2 \rangle(z)-\langle k^2 \rangle(0)$,
of the change in beam angular divergence,  rather than the
differential measure, $D$. $\triangle/\tilde I_0^2$ is shown in
Fig. 6, normalized to unity at $\tilde I_0=0.61$.

Note that we have not observed significant
departure from Gaussian $E$ fluctuations for $\tilde I_0<2$ for the parameters of figure 6,
which is consistent with the absence of self-focusing.
\begin{figure}
\begin{center}
\includegraphics[width = 3.4 in]{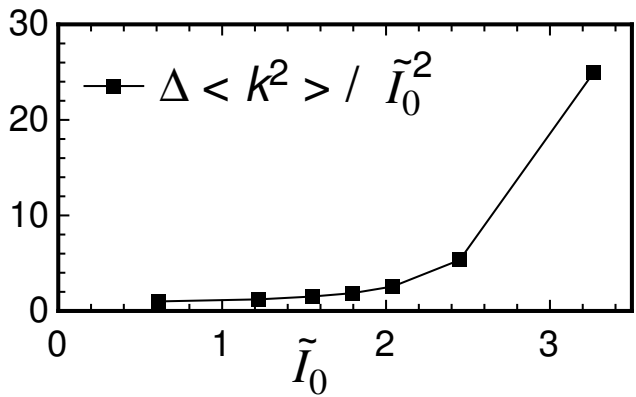}
\caption{Beam angular divergence rate increases rapidly with
$\tilde I_0$. $k_m^2z/k_0\simeq 15.7,$ $ \nu=0.0375, \tilde
T_c\simeq 0.125 .$ In contrast, Eq. $(\ref{Deq1})$ predicts a flat
curve around 1.} \label{fig:fig5}
\end{center}
\end{figure}
Therefore in this regime the effect of FSBS is benign, and perhaps
useful for NIF design purposes:  correlation lengths decrease, at
an accelerated pace compared to SSE for $\tilde I_0 \sim 1$, with $z$,
while electric field fluctuations
stay nearly Gaussian.  As a result \cite{Afeyan}, the intensity
threshold for other instabilities (e.g., backscatter SBS)
increases \cite{RoseDuBois1994}.  If $\tilde I_0>4$, there are
large non-Gaussian fluctuations of $E$, which
indicates strong self-focusing.

In conclusion, well above the FSBS threshold we observe strong self-focusing effects, while well below threshold beam propagation is  diffusive in angle
 with essential corrections to geometric optics.
In an intermediate range of intensities  the rate of angular diffusion increases with propagation.  In the weak and intermediate regimes, the diffusion results in
 decreasing correlation lengths which could be beneficial for NIF.

One of the author (P.L.) thanks E.A. Kuznetsov for helpful
discussions.

Support was provided by the Department of Energy, under contract
W-7405-ENG-36.





\begin{thebibliography}{}
\bibitem{comment3}This requires that the speed of light, $c \gg L_z / T_c$,
   where $L_z$ is the $z$ correlation length.
\bibitem{comment5}If $k$ is constrained by finite beam size effects to be, $e.g.$,  $>1/$(beam diameter),
then stability is regained for small enough $I_0$.
\bibitem{comment6}If collisonal absorption  is included in Eq. (3), with rate $\kappa$, then for $T_c\to 0$, the
 stability condition is $\lambda_{max}< \kappa$.
\end{thebibliography}
\end{document}